# Tunable band structures of polycrystalline graphene by external and mismatch strains

Jiangtao Wu · Xinghua Shi · Yujie Wei

**Abstract** Lacking a band gap largely limits the application of graphene in electronic devices. Previous study shows that grain boundaries (GBs) in polycrystalline graphene can dramatically alter the electrical properties of graphene. Here, we investigate the band structure of polycrystalline graphene tuned by externally imposed strains and intrinsic mismatch strains at the GB by density functional theory (DFT) calculations. We found that graphene with symmetrical GBs typically has zero band gap even with large uniaxial and biaxial strain. However, some particular asymmetrical GBs can open a band gap in graphene and their band structures can be substantially tuned by external strains. A maximum band gap about 0.19 eV was observed in matched-armchair GB (5, 5) | (3, 7) with a misorientation of θ=13º when the applied uniaxial strain increases to 9%. Although mismatch strain is inevitable in asymmetrical GBs, it has a small influence on the band gap of polycrystalline graphene.

## 1 Introduction

Graphene is proposed for applications in electronic devices [1-5]. The combination of super electrical transition [6-8], high thermal conductivity [9-11], tunable magnetism properties[12, 13] and other extraordinary performance [7, 14] in graphene help it to stand out while competing with conventional transistors. One critical factor limiting its applications is that large-scale pristine graphene shows semimetal property with zero band gap [5, 15]. Many strategies have been tried to realize band gap in graphene, such as cutting graphene into graphene nanoribbons (GNRs) with appropriate edge structure and width [16-21], substrate-induced gap opening method [14, 21-25] and tensile straining inducing method [26, 27]. Among those methods, opening band gap by straining is regarded as an economical and effective method in engineering practice, and there are indeed proved examples in carbon nanotube materials and Si nanowire based photonic devices [28, 29]. Recent studies also show that the presence of grain boundary (GB) can change the transport and magnetism properties of large-area graphene and open a transport gap in graphene [30-32]. Given the emerging success in synthesis for large-area polycrystalline graphene, the influence of GB defects in polycrystalline graphene on band structures is of significance. The purpose of this work is to give systematic investigations on how the presence of GBs would influence the band structure of graphene, and how externally applied strains and the mismatch strain at GBs may affect the band gap in polycrystalline graphene.

## 2 Methods

Following the rules to define the chirality of carbon nanotubes, a GB [33-35] in polycrystalline graphene can be ascribed by two translational vectors [30, 31] $\vec{V_L}(m_L, n_L)$ and $\vec{V_R}(m_R, n_R)$, which are the respective directions of the two edges of the connected grains in the GB (see Fig. 1a). In general, the GB formed by the two edges is composed of pentagon-heptagon rings. The length of the two translational vectors can be calculated as

J. Wu · X. Shi · Y. Wei (✉)
LNM, Institute of Mechanics,
Chinese Academy of Sciences,
100190 Beijing,China
email: yujie_wei@lnm.imech.ac.cn.

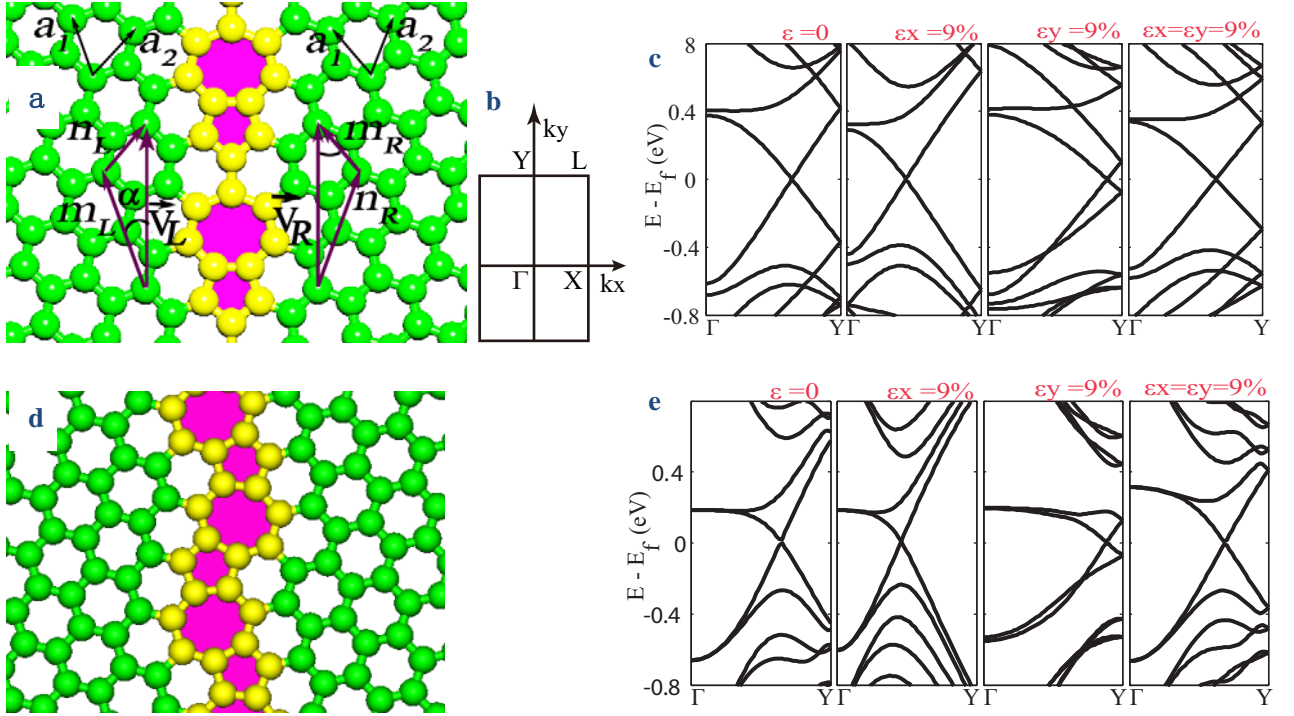

**Fig. 1** GB structure and band structure of symmetrical polycrystalline graphene. (a) Detailed GB structure of armchair tilt ($\theta=21.8°$) graphene; (b) The first Brillouin zone of all structures and special K points used to calculate the band structures (c) The corresponding band structures for the structure in (a) with different types of strains. (d) Detailed GB structure of zigzag tilted boundary with $\theta=27.8°$; (e) The corresponding band structures for the structure in (d) subjected to different strains respectively.

$$d_L = |\vec{V}_L| = \sqrt{3}a_0\sqrt{n_L^2 + n_L m_L + m_L^2}, \quad (1.a)$$

$$d_R = |\vec{V}_R| = \sqrt{3}a_0\sqrt{n_R^2 + n_R m_R + m_R^2}, \quad (1.b)$$

where $a_0$ is the C-C bond length in graphene.
The orientation angle $\alpha$ can be defined as the angle between a translational vector $\vec{V}$ and the unit vector $\vec{a}_1$, with

$$\alpha_L = \arcsin\left(\frac{\sqrt{3}n_L}{2\sqrt{n_L^2 + n_L m_L + m_L^2}}\right), \quad (2.a)$$

$$\alpha_R = \arcsin\left(\frac{\sqrt{3}n_R}{2\sqrt{n_R^2 + n_R m_R + m_R^2}}\right). \quad (2.b)$$

So the grain misorientation between the two domains can be calculated as following:

$$\theta = \arcsin\left(\frac{\sqrt{3}n_R}{2\sqrt{n_R^2 + n_R m_R + m_R^2}}\right) - \arcsin\left(\frac{\sqrt{3}n_L}{2\sqrt{n_L^2 + n_L m_L + m_L^2}}\right) \quad (3)$$

We studied the band structure of polycrystalline graphene under the influence of external strains and mismatch strains via DFT calculations. Two typical kinds of GB structures are investigated: (1) symmetrical tilt GBs ($d_L = d_R, m_L = n_R$, and $m_R = n_L$), and (2) asymmetrical GBs ($\alpha_L + \alpha_R \neq \frac{\pi}{3}$). Details about the construction of symmetrical GBs can be found in many references [36-39]. For the asymmetrical GBs, in addition to those matched-zigzag or matched-armchair GBs, we also adopted the asymmetrical GB ((5, 0) | (3, 3) GB) from Ref. 30. It is noted that matched-zigzag or matched-armchair GBs refers to the type of GBs with one side being either along zigzag or armchair direction, and adhere to a side from the opposite grain to form a GB. The final samples contain two GBs to ensure the applicability of periodic boundary conditions as discussed in Fig. S5 in Ref. 39. Uniaxial and biaxial strains are then applied by changing the unit cell. While applying a uniaxial strain to the sample, we allow the other in-plane dimension of the cell be able to relax. Each sample may subject to (a) a strain $\varepsilon_x$, i.e., applying stretch perpendicular to the GB, (b) a strain $\varepsilon_y$, applying stretch parallel to the GB, and (c) a biaxial strain with $\varepsilon_x = \varepsilon_y$.

The DFT calculations are performed with the Vienna Ab initio Simulation Package (VASP) code [40, 41]. The projector augmented wave (PAW) pseudopotentials [42, 43] and the generalized gradient approxima-tion (GGA) of the Perdew-Burke-Ernzerhof (PBE) functional [44, 45] are used. A plane-wave basis set with a kinetic-energy cut-off of 400 eV and a Monkhorst-Pack [46] k-point mesh of at least 3×5×1 (Γ icluded) are used for static electronic structure calculations. To eliminate the interactions between periodic images of graphene, a vacuum space of 20 Å along the thickness direction of graphene was used. All structures are relaxed using a conjugate gradient algorithm until the atomic forces are converged to 0.01eV/Å. The first Brillouin zone of all structures and special *K* points used to calculate the band structures are shown in Fig. 1b.

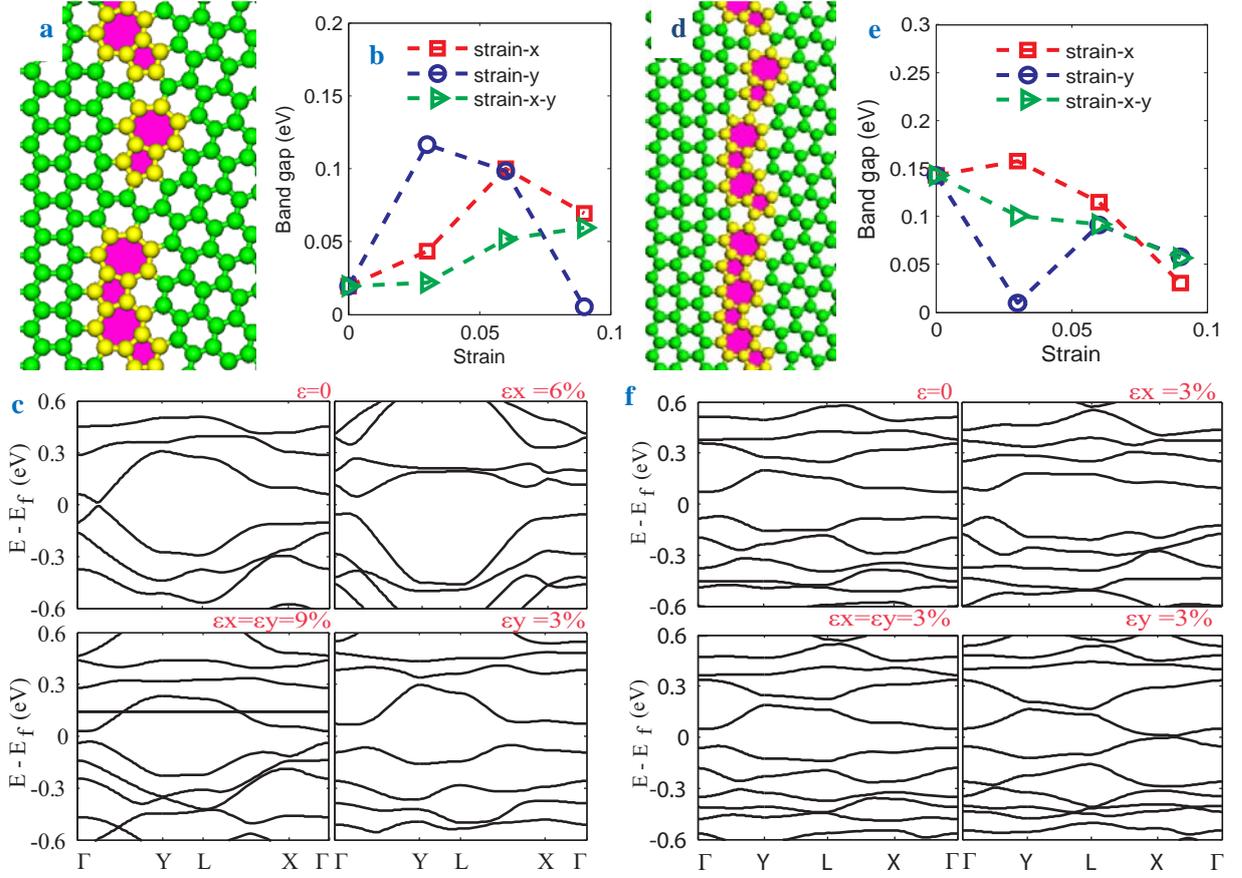

**Fig. 2** GB structure and Band structure of asymmetrical matched-zigzag-type polycrystalline graphene. (a), (d) Detailed GB structure of matched zigzag GB with misorientation of $\theta$=19.1° and $\theta$=23.4° respectively; (b), (e) The band gap for the boundary in Fig. 2(a), 2(c)as a function of strain respectively. (c), (f) The corresponding band structures of structure (a), (c) subjected to different strains respectively.

## 3  Band structures tuned by external strains

Two types of symmetrical GBs, symmetrical zigzag tilt GB and armchair tilt GB (obtained from ref. 35) are studied. The detailed structure of the symmetrical armchair tilt GB with misorientation $\theta$=21.8° after relaxation is given in Fig. 1a. Its band structures at different deformation status are given in Fig. 1c, corresponding to the relaxation status, at $\varepsilon_x$=9%, $\varepsilon_y$=9%, and $\varepsilon_x=\varepsilon_y$=9%. It is clear that this GB shows semimetal property with nearly zero band gap. Fig. 1d gives the symmetrical zigzag tilt GB with $\theta$=27.8°. We show in Figure 1e the band structures of the GB in Fig. 1d at the relaxation status, at $\varepsilon_x$=9%, $\varepsilon_y$=9%, and $\varepsilon_x=\varepsilon_y$=9%. Similar to the tilted armchair GB, the tilted zigzag GB also show semimetal property. It confirms the calculation in Ref. 32 that symmetrical GBs in graphene essentially show no band gap. This conclusion holds for those tilt GBs even with strain undulation. In addition, we find that the band gaps are insensitive to the density and arrangement of pentagon-heptagon rings in tilted GBs.

While no correlation between band gaps and applied strains is seen in symmetrical tilt GBs, the situation is different for asymmetrical GB structures. We show in Fig. 2a the structure of a matched-zigzag GB (8, 0) | (6, 3) with a misorientation of $\theta$=19.1°. A nealy zero band gap is observed in the relaxed sample, as

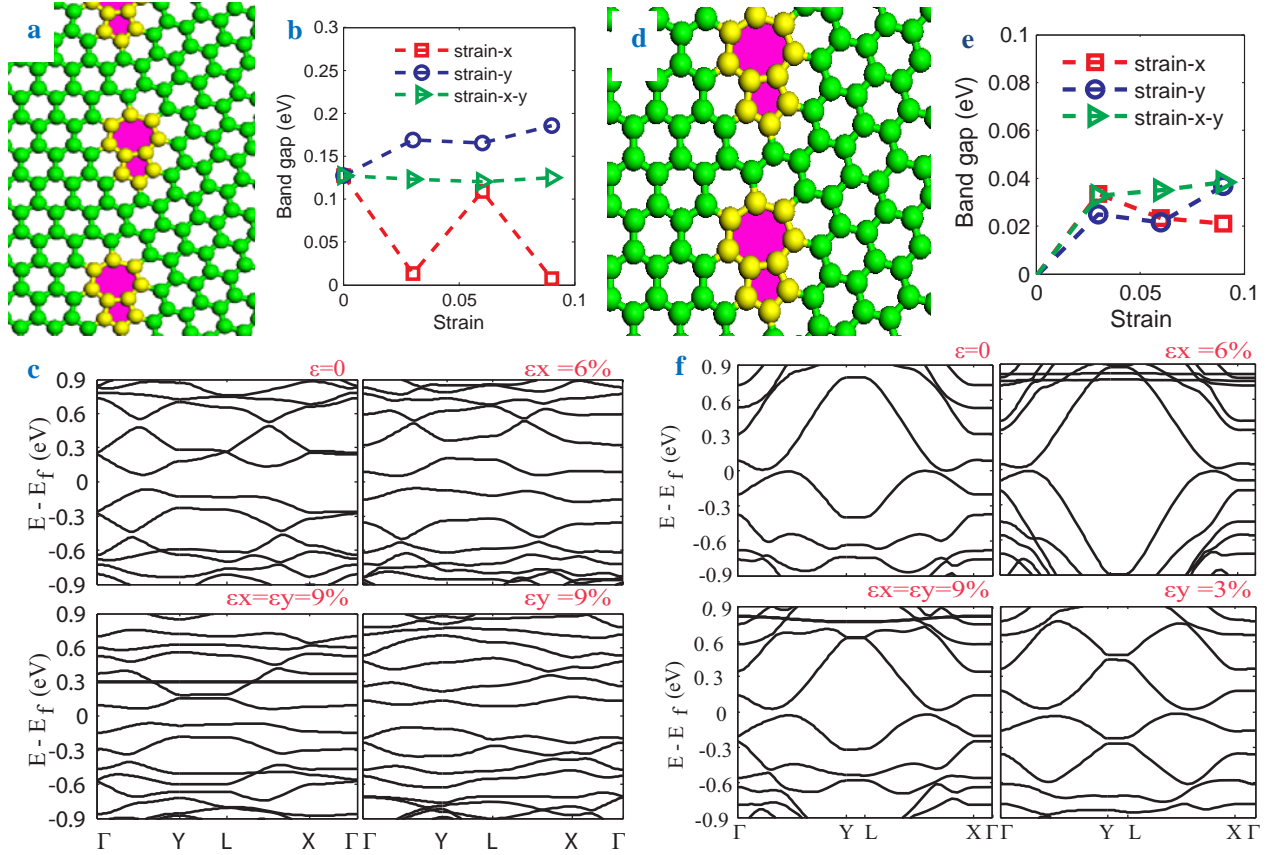

**Fig. 3** GB structure and Band structure of asymmetrical matched-armchair-type polycrystalline graphene. (a), (d) Detailed GB structure of matched armchair GB with misorientation of $\theta =13°$ and $\theta =16.1°$ respectively; (b), (e)The band gap for the boundary in Fig. 3(a), 3(b) as a function of strain respectively. (c), (f) the corresponding band structures of structure (a), (c) subjected to different strains respectively.

shown in Fig. 2c. From Fig. 2c, we find that the band gap broadens when the small strains are applied. The dependence of band gap on strains in the matched-zigzag GB is summarized in Fig. 2b. From Fig. 2b, we can find that the band gap will decrease when larger uniaxial strain is applied. The band gap becomes closed when strain uniaxial strain parallel to the GB ($\varepsilon_y$) increases to 9%. Biaxial strain can slightly broaden the width of band gap. Similarly, we show in Fig. 2d the structure of a matched-zigzag GB (13, 0) | (9, 6) with a misorientation of $\theta=23.4°$. The initial band gap in the absence of external strains is about 0.14 eV (see Fig. 2f), which differs from the same sort of GB but with $\theta=19.1°$. Fig. 2e shows the band gap of this GB as a function of strain. The band gap remains nearly no change when strain of 3% perpendicular to the GB (Fig. 3b) is applied. However, it decreases dramatically as we further increase the strain. When small uniaxial strain parallel to GB is applied, the band gap becomes closed. Increasing the biaxial strain will lead to a decrease in the band gap.

The strong dependence of band gap on applied strains in asymmetrical GBs is further explored in two matched-armchair GB with $\theta =13°$ and $\theta =16.1°$, respectively. The detailed GB structure for the matched-armchair GB with $\theta =13°$ is given in Fig. 3a. In Fig. 3c, we show in turn the band structures of the GB subjected to different strain status: after relaxation, at $\varepsilon_x =6\%$, $\varepsilon_y =9\%$, and $\varepsilon_x = \varepsilon_y =9\%$. The GB has a band gap of 0.13 eV at strain-free status. When $\varepsilon_y =9\%$ is applied, the band gap increases to 0.19eV. Strains perpendicular to a GB tend to close the band gap. As to the biaxial stain cases, the band gap does not vary much. Fig. 3b gives more comprehensive information about the band gap as a function of strain when different types of strains are applied to the GB. For the matched-armchair GB with $\theta =16.1°$ (Fig. 3d), there is no band gap at the strain free status (Fig. 3f). Unlike the other matched-armchair GB with $\theta =13°$, the band gap of the GB at different strains (Fig. 3f) shows almost no changes in response to the applied strains. The curves for band gap versus strain for the matched-armchair GB (Fig. 3d) are shown in Fig. 3e. It is found that the band gaps are neither sensitive to the magnitude of strains, nor to the type of the applied strain.

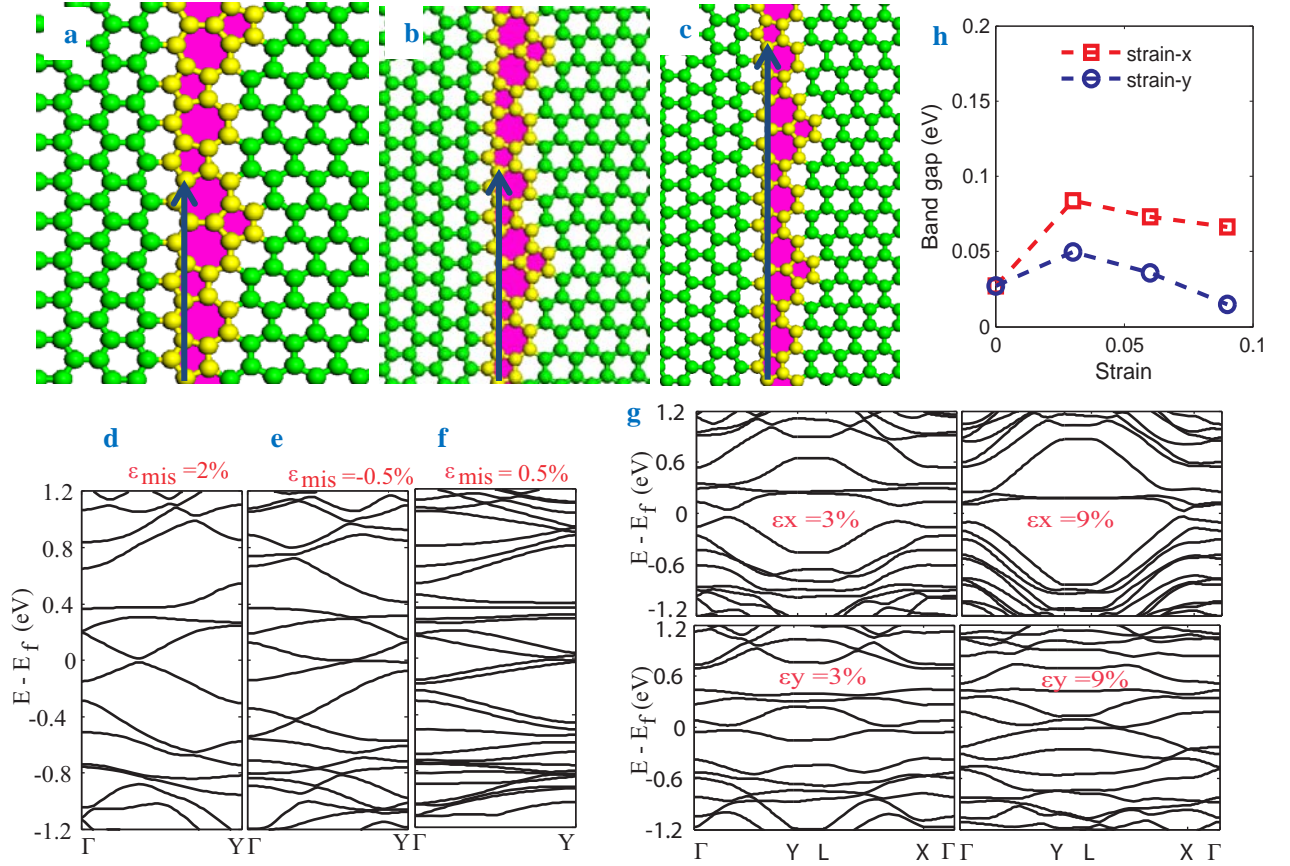

**Fig. 4** Atomic structures and band structures of asymmetrical GBs with the same misorientation of θ =30°. Detailed atomic structures for different GBs: (a) (0, 5) | (3, 3), (b) (0, 7) | (4, 4), and (c) (0, 12) | (7, 7). (d), (e) and (f) are the band structures for (a), (b) and (c), respectively. (g) The band structures of the (0, 5)| (3, 3) boundary subjected to different tensile strain. (h) The band gap of boundary in Fig. 4(a) as a function of applied uniaxial strain.

**4 Band structures influenced by mismatch strains**

So far, we have demonstrated that the global strain can significantly influence the band gap of asymmetrical GBs in polycrystalline graphene but no perceivable effects to symmetrical tilt GBs. Now we focus on understanding the influence of mismatch strains in GBs to the band structures in polycrystalline graphene. As two mismatched lattices adhering together to form a GB, mismatch strains arise naturally in the GB. The mismatch strain in an asymmetrical GB in graphene can be defined as $\varepsilon_{mis} = \frac{d_R - d_L}{d_R + d_L}$. To investigate the influence of mismatch strain on band structures in graphene, we constructed three GB structures with the same misorientation of θ =30°. Fig. 6(a), 6(b) and 6(c) show the detailed structures of (0, 5) | (3, 3), (0, 7) | (4, 4), and (0, 12) | (7, 7) GBs respectively. Their corresponding mismatch strain is about 2%, -0.5% and 0.5%, respectively. Fig. 4d to 4f show respectively the band structures of the three GBs with different mismatch strains. These calculations indicate that the band gap does not open much with mismatched strain. The influence of external strains on band structures in GBs with initial mismatch strain is also investigated. We applied uniaxial tensile strain to the (0, 5) | (3, 3) GB. Fig. 4g shows the band structures of the GB when the strains are normal and parallel to the GB. Fig. 4h shows the variation of band gaps with applied uniaxial strain. Clearly, the band gap can be tuned by uniaxial tensile strains.

**4 Conclusion**

In summary, we have investigated the band structure of symmetrical and asymmetrical GB structures of graphene, as well as band structures tuned by externally applied strains or internal mismatch strains. For symmetrical GBs in graphene, there is essentially no band gap even with strain undulation, regardless the density GBs. For some particular asymmetrical GBs, their band structures and band gaps can be substantially tuned by applying external strains. From our DFT calculations, the band gap in some particular asymmetrical GBs can be about 0.1- 0.2 eV when particular strains are applied. This band gap is not a wide band gap in contrast to that of silicon, which is about 1.1eV. However, we note that the DFT calculation is performed with generalized gradient approximation (GGA) functional within Kohn-Sham formalism. It is well recognized that such DFT calculations typically underestimate the band gap of semiconductors and insulators [47-49]. We expect a reliable yet large band gap if other improved methods with more accurate band gap prediction such as GW approximation and HSE calculation could be employed. Given all our calculations involves a system with 170 to 414 atoms, such calculations are still computationally too expensive and are beyond our capability at this moment.

**Acknowledgment**

The authors acknowledge support from National Natural Science Foundation of China (NSFC) (11021262) and MOST 973 of China (Nr. 2012CB937500) for Y.W., and NSFC (11023001) for X.S. Computation is supported by the Supercomputing Center of Chinese Academy of Sciences.